\documentstyle[12pt]{article}

\begin{document}
\baselineskip=1.5\baselineskip

\vspace{0.1cm}
\begin{center}
{\Large\bf
Exact supersymmetry in the nonrelativistic hydrogen atom }
\end{center}
\vspace {5mm}
\rm
\begin{center}
R.\ D.\ Tangerman\\
{\small\it  National Institute for Nuclear Physics and High Energy Physics
(NIKHEF-K), \\
P.O. Box 41882, 1009 DB Amsterdam, The Netherlands}\\
 and \\
 J. A. Tjon \\
 {\small\it Institute for Theoretical Physics,\\
 Princetonplein 5,
 P.O. Box 80.006,
 3508 TA Utrecht,
 The Netherlands}
\end{center}
\vspace {1.0cm}

\begin{abstract}
We consider a Pauli particle in a Coulomb field.
The supersymmetric Hamiltonian is constructed,
by explicitly giving the two supercharges $Q_{1}$ and $  Q_{2}$
in the full three-dimensional space and which
together with the Hamiltonian, are shown to constitute an $S(2)$
superalgebra.  This offers an alternative way of grouping the
energy eigenstates into irreducible representations of this
symmetry group of the Hamiltonian.
\end{abstract}

\vspace{1cm}

\newpage

\section{Introduction}

In recent years considerable attention has been paid to super\-sym\-me\-tric
\linebreak (SUSY) quantum mechanics (for a review see ~\cite{Lah}
and the references therein).
An interesting suggestion in this context has been made by Kosteleck\'{y}
and Nieto~\cite{Kos},
that SUSY may be realized in atomic systems as a good symmetry to
characterize the spectrum of alkali-metal atoms.
In the discussion of SUSY quantum mechanics one usually considers
one-dimensional systems. For the simplest case (N=2) the SUSY
Hamiltonian is expressed in terms of two operators $Q_{1}$ and
$  Q_{2}$, called supercharges, obeying the S(2) product rules
\begin{eqnarray}
\label{twee}
& & [  Q_{1},\tilde{H}]=[  Q_{2},\tilde{H}]=0,
 \nonumber \\
& &\{  Q_{1},  Q_{2}\}=0, \\
& &\{  Q_{1},  Q_{1}\}=\{  Q_{2},  Q_{2}\}=\tilde{H},\nonumber
\end{eqnarray}
with $\tilde{H}=f(H)$ being an
invertable function of the original Hamiltonian $H$.

Applying it to the case of the hydrogen atom~\cite{Kos,Kos1},
the corresponding one-dimensional superpotentials~\cite{Suku}
can be determined by considering the radial
part of the Schr\"{o}dinger equation.
In higher dimensions superalgebras can explain accidental spectrum
degeneracies as has been shown for the three-dimensional
harmonic oscillator~\cite{Bal}. D'Hoker and Vinet have discovered a
supersymmetry for a \mbox{spin-$\frac{1}{2}$} particle in the presence of a
dyon field~\cite{Hok}. However, for this they need the introduction of an
additional particle.
For the hydrogen problem it is expected that an explicit form of
a SUSY Hamiltonian in three dimensions does exist. We indeed
show in this paper how the supercharges can be obtained in this
case, by just using the spin degrees of freedom of the electron. We stress the
fact that no extra particles are needed in our formalism.

In the next section we construct a $N=2$ supersymmetry for the
Coulomb potential. Use is made of the well-known $SO(4)$ symmetry.
The multiplets are identified in section 3 and the
connection is made with radial supersymmetries.
Finally, in section 4 we discuss the possible implications for other systems,
in particular the relativistic \mbox{spin-$\frac{1}{2}$} particle in a Coulomb
field.

\section{Construction of the charges}

The accidental degeneracies of the hydrogen atom bound state spectrum
can be ascribed to an $SO(4)$ symmetry~\cite{Pau} of its Hamiltonian
\begin{equation}\label{coulomb}
 H=\mbox{$\frac{1}{2}$}{\bf p}^{2} - \frac{1}{r}. \end{equation}
Apart from the obvious
$SO(3)$-invariance, generated by the angular momenta \[{\bf
L}=\mbox{$\frac{1}{2}$}\left({\bf x}\times{\bf p}- {\bf p}\times{\bf
x}\right),\]
it has three additional constants of the motion, which together form the
Runge-Lenz vector \begin{equation}{\bf A}_{0}=\mbox{$\frac{1}{2}$}
\left({\bf p}\times{\bf L}-{\bf L}\times{\bf p}\right)-\frac{{\bf x}}{r}.
\end{equation}
Only after proper normalization
\begin{equation}
 {\bf A} \equiv \left\{ \begin{array}{lr}
	\frac{{\bf A}_{0}}{\sqrt{-2H}} & \mbox{if $E<0$} \\ {\bf A}_{0}
	& \mbox{if $E=0$} \\ \frac{{\bf A}_{0}}{\sqrt{2H}} & \mbox{if
	$E>0$} \end{array} \right. , \end{equation}
one obtains a closed algebra. For the bound states ($E<0$) the product rules
are
\begin{equation}\label{algebra} \begin{array}{l}  [ L_{i},L_{j}
]=i\varepsilon_{ijk}L_{k},\\
		    {[} L_{i},A_{j} ]=i\varepsilon_{ijk}A_{k},\\ {[}
		    A_{i},A_{j} ]=i\varepsilon_{ijk}L_{k}.
  \end{array} \end{equation}
  They form the defining relations of an $SO(4)$
algebra\footnote{In contrast, the scattering states $(E>0)$ are
$SO(3,1)$-multiplets.}. The two Casimir operators
\begin{eqnarray}\label{Casimir1}
&C_{1}= {\bf L}^{2}+{\bf A}^{2},&\\& C_{2}= {\bf A}\cdot{\bf L},&
\label{Casimir2} \end{eqnarray}
in the present realization assume the form
\begin{eqnarray}
&C_{1}=-\frac{1}{2H}-1,&\\ &C_{2}=0.& \end{eqnarray} The general versions of
these
relations, being valid at {\em all} energies, are
\begin{eqnarray}
&{\bf A}_{0}^{2}=2H({\bf L}^{2}+1) +1, &\label{achttien}\\ & {\bf
A}_{0}\cdot{\bf L}={\bf L}\cdot{\bf A}_{0}=0.&\label{zeventien}
\end{eqnarray}
To construct the supercharges for the Coulomb problem we
also include the spin degrees of freedom. As a consequence the
symmetry algebra is enlarged with the $SU(2)$ generators
${\bf S}=\mbox{$\frac{1}{2}$}{\bf \sigma}$, resulting into a
$SO(4)\times SU(2)$ symmetry group for the bound states.
Usually one chooses to diagonalize the set of observables $H$, ${\bf
J}^{2}$, ${\bf L}^{2}$ and $J_{3}$.
Following Biedenharn and Louck~\cite{Bie},
the two operators  ${\bf J}^{2}$
and ${\bf L}^{2}$ can be replaced by a single scalar $\cal K$, defined
by
\begin{equation}
{\cal K}\equiv -(2{\bf S}\cdot{\bf L}+1).
\end{equation}
We have
\begin{eqnarray}
\left\{\begin{array}{l} {\bf L}^{2}={\cal K}({\cal
K}+1)\\ {\bf J}^{2}={\cal K}^2-1\end{array}\right.&
\Rightarrow&\left\{\begin{array}{l} \ell(\kappa)= |\kappa
|+\mbox{$\frac{1}{2}$}
( \mbox{sgn}\kappa -1)\\ j(\kappa)=|\kappa
|-\mbox{$\frac{1}{2}$}\end{array}\right. ,
\end{eqnarray}
where $\kappa$ takes
the values $\{\pm 1,\pm 2,\ldots\}$. It is clear that states with fixed
$j$ and $\ell=j\pm\mbox{$\frac{1}{2}$}$ correspond to equal $|\kappa|$ but with
$ \mbox{sgn}\kappa=\pm 1$.

A $Z_2$ -grading can now be introduced in the Hilbert space of
states by classifying the states to be
even or odd with respect to the parity operator
\[ P_{\kappa}\equiv \frac{{\cal K}}{|\kappa|},\]
i.e. having eigenvalues $\pm  \mbox{sgn}\kappa$ respectively.
Equally, linear operators can be
assigned a grade. An operator is even, if it commutes with
$P_{\kappa}$, whereas operators anti-commuting with $P_{\kappa}$ are
called odd.
As an example of an even operator we mention $H$, following from the
fact that $\cal K$ is built from symmetry operators of $H$. Of course,
$\cal K$ itself is even as well. To find odd operators, we make use of
the following theorem.\\[5mm] {\bf Theorem.} {\it Suppose ${\bf V}$ is a
vector with respect to the orbital angular momentum ${\bf L}$ that is
also perpendicular to ${\bf L}$ \begin{eqnarray}
&&[L_{i},V_{j}]=i\varepsilon_{ijk}V_{k},  \label{vector}\\ &&{\bf
L}\cdot{\bf V}={\bf V}\cdot{\bf L}=0, \end{eqnarray} then ${\cal K}$ {\em
anticommutes} with the $\bf J$-scalar ${\bf S}\cdot{\bf V}$.}\\[5mm]
For a proof see Ref.~\cite{Bie}. This theorem
supplies us with odd operators  ${\bf S}\cdot{\bf V}$,
with ${\bf V}$ equal to for example
${\bf p}$, ${\bf r}$ or ${\bf A}_{0}$.

Considering the square of the odd operator  ${\bf S}\cdot{\bf A}_{0}$,
it is readily shown that
\begin{equation}
\label{lading} ({\bf S}\cdot{\bf A}_{0})^{2}=\mbox{$\frac{1}{2}$}
H{\cal K}^{2}+\mbox{$\frac{1}{4}$},
\end{equation}
where Eq.~(\ref{achttien}) has been used. Restricting ourself to
the subspace ${\cal H}_{k}$ of fixed $|\kappa|\equiv k$, this
can be rewritten as
\begin{equation}\label{kern}
2\left(\frac{{\bf S}\cdot{\bf
A}_{0}}{k}\right)^{2}=H+\frac{1}{2k^{2}}.
\end{equation}
We now define the supercharge
\[  Q_{1}\equiv\frac{{\bf S}\cdot{\bf A}_{0}}{k},\]
and the shifted Hamiltonian
\[\tilde{H}\equiv H+\frac{1}{2k^{2}}.\]
Eq.~(\ref{kern}) can then be identified with the $S(2)$-product
rule $\{  Q_{1},  Q_{1}\}=\tilde{H}$.
Moreover, using the fact that $  Q_{1}$ is odd and $H$
is even, we readily identify the second charge as
\[  Q_{2}\equiv
i  Q_{1} P_{\kappa}=\frac{i}{k^{2}}({\bf S}\cdot{\bf A}_{0}){\cal K},\]
thereby completing the $S(2)$ symmetry algebra.

Instead of working in the Hermitian representation of $S(2)$ as
described
above, one often uses the odd ladder operators\begin{equation}\label{step}
Q_{\pm}\equiv\frac{1}{\sqrt{2}}(  Q_{1}\pm i  Q_{2})=\frac{1}{\sqrt{2}k}({\bf
S}\cdot{\bf A}_{0})(1\mp P_{\kappa}). \end{equation}Note that for the even
states
$Q_{+}=0$ and $Q_{-}=\frac{\sqrt{2}{\bf S}\cdot{\bf A}_{0}}{k}$,
whereas for the odd states their roles are reversed.

\section{The hydrogen spectrum revisited}

We now turn to discuss the implications of the constructed
$S(2)$ symmetry on the spectrum.
The consequence
of a symmetry group of a Hamiltonian is that the energy eigenspaces
consist of irreducible representations of it.
The irreducible representations of $S(2)$ are
known to be either one- or two-dimensional.
Consider the energy eigenstates of the Pauli particle, given by
\begin{equation}\label{decomposition}
\phi_{E\ell(\kappa) j(\kappa) m}(r,\theta,\varphi)
=R_{E\ell(\kappa)}(r)Y^{[\ell(\kappa)\mbox{$\frac{1}{2}$}]j(\kappa)m}(\theta,\varphi),
\end{equation}
where $Y^{[\ell(\kappa)\mbox{$\frac{1}{2}$}]j(\kappa)m}\equiv\chi_{m}^{\kappa}$
are
the Pauli central field spinors. The quantum number
$\kappa$ assumes the values $\pm k$.
Applying the ladder operators~(\ref{step}) to
Eq.~(\ref{decomposition}) yields either zero or
$\frac{\sqrt{2}{\bf S}\cdot{\bf A}_{0}}{k}$.
Moreover, since
$\{{\cal K},{\bf S}\cdot{\bf \hat{r}}\}=0$ (see theorem)
and $({\bf S}\cdot{\bf \hat{r}})^{2}=\mbox{$\frac{1}{4}$}$,
we obtain the nice property that
\begin{equation}\label{anactie}
({\bf S}\cdot{\bf\hat{r}})\chi_{m}^{\pm
k}=-\mbox{$\frac{1}{2}$}\chi_{m}^{\mp k}.
\end{equation}
Using
Eq.~(\ref{kern}) and appropriate phase conventions,
we find
\begin{equation}\label{actie}
\left( \frac{\sqrt{2}{\bf S}\cdot{\bf A}_{0}}{k}
\right) \phi_{E\,\pm k\,
m}=-\left(E+\frac{1}{2k^{2}}\right)^{\mbox{$\frac{1}{2}$}}\phi_{E\,\mp k\,m}.
\end{equation}
Hence states within the subspace ${\cal H}_{k}$, with fixed $E$ and $m$,
transform irreducibly under the superalgebra $S(2)$. The multiplets
have dimensionality two, unless $E=-\frac{1}{2k^{2}}$, in which case
they are one-dimensional.

Eq.~(\ref{actie}) can now be reduced to a set of radial
equations.  Using the identity
\begin{equation}\label{factor}
{\bf S}\cdot{\bf A}_{0}=\left[ 1-{\cal
K}\left( i{\bf \hat{r}}\cdot{\bf p} +\frac{{\cal
K}+1}{r}\right)\right]({\bf S}\cdot{\bf \hat{r}}).
\end{equation}
and substituting Eqs.~( \ref{anactie}, \ref{factor})
in Eq.~(\ref{actie}) we find
($\kappa=\pm k$)
\begin{equation}
\frac{1}{\sqrt{2}k}\left[ 1-\kappa\left( \frac{d}{dr}+\frac{\kappa
+1}{r}\right)\right]R_{E\ell(\kappa)}(r)=\left(
E+\frac{1}{2k^{2}}\right)^{\mbox{$\frac{1}{2}$}}R_{E\ell(-\kappa)}(r).
\end{equation}
Introducing furthermore
\( rR_{E\ell}(r)=\chi_{E\ell}(r) \) and
identifying $k=\ell +1$, these two equations can be written
as
\[ \left(\begin{array}{cc}
0&A^{-}(\ell)\\A^{+}(\ell)&0\end{array}\right) \left(\begin{array}{c}
\chi_{E\ell}(r)\\\chi_{E\ell+1}(r)\end{array}\right)
=\sqrt{E+\frac{1}{2(\ell+1)^{2}}} \left(\begin{array}{c}
\chi_{E\ell}(r)\\\chi_{E\ell+1}(r)\end{array}\right) ,\]
where
\[
A^{\pm}(\ell)= \frac{1}{\sqrt{2}}\left[ \pm \frac{d}{dr}-\frac{\ell +1}{r}
+ \frac{1}{\ell +1}\right] .
\]
The radial
ladder operators $A^{\pm}(\ell)$ are identical to those found in
the radial SUSY studies of the Coulomb problem~\cite{Kos,Ama}. We
stress the fact that in those cases the angular momentum is not really
affected, because the supercharges are purely radial.  In our
three-dimensional study the $\ell$-value {\em is} changed by the
charges, as is demonstrated by Eq.~(\ref{anactie}).

We are now in a position to reinterpret the spectrum of a
nonrelativistic \mbox{spin-$\mbox{$\frac{1}{2}$}$} particle in a Coulomb
field~(see
Fig.~ 1).

\begin{description}
\item[-]The degeneracies of levels with fixed $j$
and $m$ but with $\ell=j\pm\mbox{$\frac{1}{2}$}$ are a consequence of the
$S(2)$
supersymmetry algebra, constructed above. The degeneracy level is two,
except for the states with $\ell=j-\mbox{$\frac{1}{2}$}$ and
$E=-\frac{1}{2(\ell+1)^{2}}$, which are non-degenerate.
\item[-]The
degeneracies of fixed $\ell$ and $m$ but $j=\ell\pm\mbox{$\frac{1}{2}$}$ are
due to
kinematic independence of the electron spin in the nonrelativistic
regime. Ladder operators between these states can be constructed out
of the vector operator ${\bf S}\times{\bf L}$ (see e.g.~\cite{Bru}).
\item[-]The $(2j+1)$-fold degeneracy of states with fixed $j$ and
$\ell$ is due to rotational invariance of the interaction.
\end{description}

\section{Concluding remarks}

In this paper we have used the
operators ${\bf A}_{0}$, ${\bf L}$ and ${\bf S}$ to
construct the charges $  Q_{1}$ and $  Q_{2}$ of an $S(2)$ supersymmetry
algebra of the nonrelativistic Coulomb problem. This supersymmetry
occurs for {\em all} levels with fixed $j$ and $m$, as opposed to the
$SO(4)$, which only concerns the bound states.
The results derived in this paper, show that supersymmetry
in the Coulomb problem can be extended  beyond
the one-dimensional formulation. It is not clear in
general in
how far this also applies to other physical systems, where
supersymmetry is expected to hold. It is therefore of
interest to study also these other cases. In particular, it
may offer a possibility to explore the case of a \mbox{spin-$\frac{1}{2}$}
particle in a dyon field. The symmetry and spectrum-generating algebra of such
a system are identical to those of the hydrogen atom~\cite{Vin}. It therefore
seems reasonable to expect the $S(2)$ supersymmetry to hold as well.

As another object of interest, we mention the Dirac particle in a Coulomb
field. The relativistic spin-orbit coupling breaks the degeneracies of equal
$\ell$. However, as can
be seen in Fig.~2, the relativistic
hydrogen spectrum still exhibits the features of the $S(2)$
supersymmetry of equal $j$. Although this was also
noticed by Sukumar~\cite{Suk},
the corresponding supercharges have not yet been constructed.
Our study strongly suggests that the symmetry group of the
relativistic Coulomb problem is $SO(3)\times
S(2)$, being the remainder of the $SO(4)\times SU(2)$ symmetry
of the nonrelativistic problem. Clearly, it should be
interesting to also find the explicit form of the supercharges
in this case. If the Lamb shift, due to vacuum fluctuations,
is taken into account, the $S(2)$ symmetry is broken
as well, leaving only the rotational $SO(3)$.

\newpage
\noindent
{\bf \Large Figure Captions.}
\vspace{0.5cm}

\noindent
Fig. 1 The nonrelativistic hydrogen spectrum. All levels of equal $j$ and $m$
(including those in the continuum) are connected by an $S(2)$ supersymmetry.
\vspace{0.5cm}

\noindent
Fig. 2
The characteristic feature of $S(2)$ supersymmetry in
the relativistic hydrogen spectrum (not to scale).

\newpage
\setlength{\unitlength}{7mm}
\begin{figure}[htb] \begin{picture}(15.5,12) \thinlines
\put(2,0){\line(0,1){11.5}} \put(2,1){\makebox(0,0)[r]{$-13.6-$}}
\put(2,11){\makebox(0,0)[r]{$0.0-$}}
\put(1,5){\makebox(0,0)[t]{$\stackrel{E}{(eV)}$}}
\put(1,5.3){\vector(0,1){1.5}}
\multiput(4,0.4)(7,0){2}{\line(1,0){3.5}}
\multiput(4,0.4)(3.5,0){2}{\line(0,1){0.2}}
\multiput(11,0.4)(3.5,0){2}{\line(0,1){0.2}}
\put(5.75,0.4){\makebox(0,0)[t]{$j=\mbox{$\frac{1}{2}$}$}}
\put(12.75,0.4){\makebox(0,0)[t]{$j=\frac{3}{2}$}}
\put(4,0.6){\makebox(0,0)[b]{$\ell=0$}}
\put(7.5,0.65){\makebox(0,0)[b]{$\ell=1$}}
\put(11,0.65){\makebox(0,0)[b]{$\ell=1$}}
\put(14.5,0.65){\makebox(0,0)[b]{$\ell=2$}} \thicklines
\put(3,1){\line(1,0){2}} \multiput(3,8.5)(3.5,0){3}{\line(1,0){2}}
\multiput(3,9.89)(3.5,0){4}{\line(1,0){2}}
\multiput(3,10.38)(3.5,0){4}{\line(1,0){2}}
\multiput(3,10.6)(3.5,0){4}{\line(1,0){2}}
\multiput(3,10.72)(3.5,0){4}{\line(1,0){2}}
\multiput(3,11)(3.5,0){4}{\makebox(2,1)[bl]{\rule{14mm}{5mm}}}
 \end{picture}
\end{figure}
\vspace{5cm}
\noindent
\begin{center}
{\bf Figure 1.}
\end{center}
\newpage
\setlength{\unitlength}{7mm}
\begin{figure}[htb] \begin{picture}(15.5,12) \thinlines
\put(2,0){\line(0,1){11.5}}\put(1,5){\makebox(0,0)[t]{$E$}}
\put(1,5.3){\vector(0,1){1.5}}
\multiput(4,0.4)(7,0){2}{\line(1,0){3.5}}
\multiput(4,0.4)(3.5,0){2}{\line(0,1){0.2}}
\multiput(11,0.4)(3.5,0){2}{\line(0,1){0.2}}
\put(5.75,0.4){\makebox(0,0)[t]{$j=\mbox{$\frac{1}{2}$}$}}
\put(12.75,0.4){\makebox(0,0)[t]{$j=\frac{3}{2}$}}
\put(4,0.6){\makebox(0,0)[b]{$\kappa=-1$}}
\put(7.5,0.65){\makebox(0,0)[b]{$\kappa=1$}}
\put(11,0.65){\makebox(0,0)[b]{$\kappa=-2$}}
\put(14.5,0.65){\makebox(0,0)[b]{$\kappa=2$}} \thicklines
\put(3,1){\line(1,0){2}} \multiput(3,8.5)(3.5,0){2}{\line(1,0){2}}
\put(10,8.8){\line(1,0){2}}
\multiput(3,9.89)(3.5,0){2}{\line(1,0){2}}
\multiput(10,10.1)(3.5,0){2}{\line(1,0){2}}
\multiput(3,10.38)(3.5,0){2}{\line(1,0){2}}
\multiput(10,10.5)(3.5,0){2}{\line(1,0){2}}
\end{picture}
\end{figure}
\vspace{5cm}
\noindent
\begin{center}
{\bf Figure 2.}
\end{center}

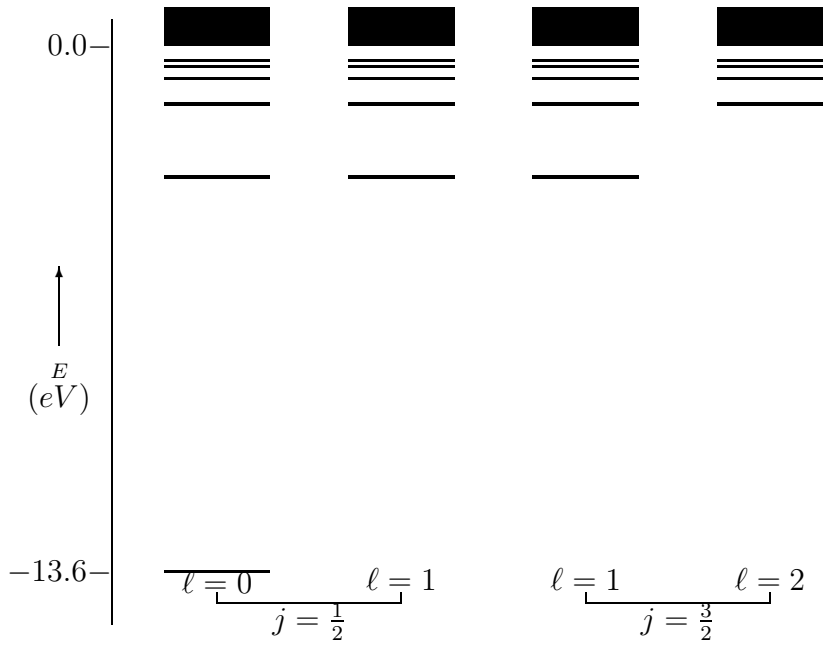
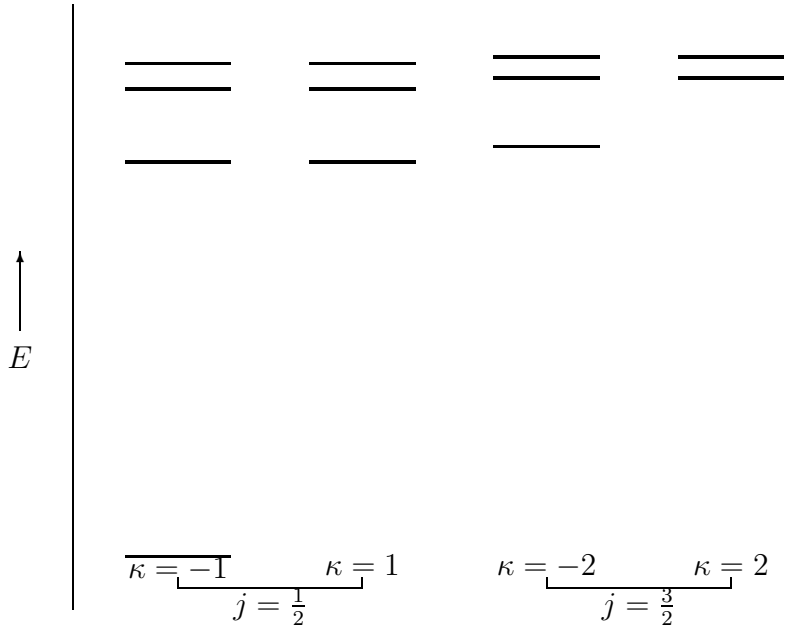
\end{document}